%% file: main.tex
\documentclass[sigconf,authorversion,nonacm,balance=false]{acmart}

\usepackage{popets}
\setcopyright{popets}
\copyrightyear{YYYY}
\acmYear{YYYY}
\acmVolume{YYYY}
\acmNumber{X}
\acmDOI{XXXXXXX.XXXXXXX}
\acmISBN{}
\acmConference{}
\input{imports.tex}

\begin{document}

\title[]{Systematic Evaluation of Geolocation Privacy Mechanisms}


\author{Alban Héon$^*$, Ryan Sheatsley$^\mathsection$, Quinn Burke$^\mathsection$, Blaine Hoak$^\mathsection$, Eric Pauley$^\mathsection$,\\ Yohan Beugin$^\mathsection$, Patrick McDaniel$^\mathsection$}
\affiliation{%
\institution{The Pennsylvania State University$^*$, University of Wisconsin-Madison$^\mathsection$}
\city{}
\state{}
\country{}
}
\email{aheon@psu.edu,sheatsley@wisc.edu, [qkb,bhoak,epauley,ybeugin,mcdaniel]@cs.wisc.edu}


\renewcommand{\shortauthors}{Héon et al.}

\begin{abstract}
\input{00.abstract}

\end{abstract}


\maketitle{}

\input{01.introduction}

\input{02.background}
\input{03.methodology}

\input{04.evaluation}
\input{05.discussion}
\input{06.conclusion}

\begin{acks}
\textbf{Funding acknowledgment:}
This material is based upon work supported by the National Science Foundation
under Grant No's. CNS-1805310 and CNS-1900873. Any opinions, findings, and
conclusions or recommendations expressed in this material are those of the
author(s) and do not necessarily reflect the views of the National Science
Foundation.
\end{acks}

\bibliographystyle{ACM-Reference-Format}
\bibliography{refs}

\appendix{}
\input{A.appendix-a}

\end{document}

%% file: imports.tex
\usepackage{placeins}

\usepackage{enumitem}
\usepackage{tcolorbox}
\usepackage[all]{nowidow}
\usepackage{siunitx}
\usepackage[linesnumbered,ruled,vlined]{algorithm2e}
\usepackage{multirow}
\usepackage{amsmath}
\usepackage{caption}
\usepackage{subcaption}
\usepackage{colortbl}

\usepackage{diagbox, eqparbox, hhline}
\newcommand{\eg}{e.g.,}
\newcommand{\ie}{i.e.,}

%% file: 00.abstract.tex

Location data privacy has become a serious concern for users as Location Based
Services (LBSs) have become an important part of their life. It is possible for
malicious parties having access to geolocation data to learn sensitive
information about the user such as religion or political views. Location Privacy
Preserving Mechanisms (LPPMs) have been proposed by previous works to ensure the
privacy of the shared data while allowing the users to use LBSs. But there is no
clear view of which mechanism to use according to the scenario in which the user
makes use of a LBS. The scenario is the way the user is using a LBS (frequency
of reports, number of reports). In this paper, we study the sensitivity of LPPMs
on the scenario on which they are used. We propose a framework to systematically
evaluate LPPMs by considering an exhaustive combination of LPPMs, attacks and
metrics. Using our framework we compare a selection of LPPMs including an
improved mechanism that we introduce. By evaluating over a variety of scenarios,
we find that the efficacy (privacy, utility, and robustness) of the studied
mechanisms is dependent on the scenario: for example the privacy of Planar
Laplace geo-indistinguishability is greatly reduced in a continuous scenario. We
show that the scenario is essential to consider when choosing an obfuscation
mechanism for a given application.

%% file: 01.introduction.tex
\section{Introduction}\label{introduction}

Privacy could be defined as the right for the user to choose what happens to
their personal data: who can access it and to what
extent~\cite{westin_privacy_1970}. The problem with location data is that even
if the user agrees to share it with some Location Based Service (LBS), the
shared data could be used by the LBS to gain insight on some private data the
user did not explicitly agree to share. It has been shown that location data
could be leveraged to learn private information on the user such as their home
or work address, habits, sexual preferences, religion, or political
views~\cite{cox_location_2022,thompson_opinion_2019,cyphers_how_2022,hern_fitness_2018,eaton_catholics_2019}.

Privacy concerns for geolocation data have led to the development of Location
Privacy Preserving Mechanisms (LPPMs). Those mechanisms aim at protecting the
privacy of the users regarding their location data while allowing them to make
use of LBSs. Developing a new LPPM has always been a compromise between the
utility the user can get from the data after applying the mechanism and the
privacy ensured by the mechanism~\cite{zhang_online_2019}. When using a
mechanism that adds noise to ensure the privacy, the noise will also reduce the
utility of the data because it will be less precise.

Previous works have introduced new LPPMs, but it is unclear which should be used
rather than the other ones when the nature of the location disclosure differs:
\ie{} when users report their location to LBS frequently or not, or for a long
period of time or not.

Looking at the literature, we identify three dimensions that characterize the
performance of a mechanism: privacy, robustness, and utility. Privacy is the
inverse of the amount of information that we can get from the data in the
absence of an adversary: high privacy means that the data does not convey much
information. Robustness is the level of privacy that a mechanism maintains in
the presence of an adversary. Utility is the degree to which the data is useful.

In this paper, we study the influence of the scenario on the performance across
multiple mechanisms. To this end, we propose a framework to help us
systematically evaluate LPPMs. The framework allows us to run the same mechanism
in a variety of scenarios for multiple combinations of LPPMs, attacks, and
metrics.

Our framework is made of four parts: the datasets, the LPPMs, the attacks, and
the metrics. The choice for this decomposition was made to allow a systematic
evaluation of the mechanisms and their dependence on the scenario on which they
are used. Being able to measure the performance of the mechanisms for multiple
datasets will allow to observe their sensitivity to the scenario. Being able to
evaluate for different attacks will give us insight on the performance and the
robustness of the mechanisms.

Using our framework we compare a selection of LPPMs including an improved
mechanism that we introduce. We evaluate over a variety of scenarios: four
different datasets and then one dataset that we sub-sample to study further the
sensitivity of the mechanisms to the scenario of the datasets. We ask the
following questions: is there a most private mechanism, does frequency of report
affect the mechanisms, and does distance between two consecutive points affect
the mechanisms?

We find that the performance that obfuscation mechanisms provide is highly
dependent on the scenario on which they are used. For example, we find that the
privacy of Planar Laplace geo-indistinguishability is greatly reduced in a
continuous scenario (20\% points of interest recall) compared to a sparse
scenario (45\% points of interest recall).

The contributions of this work are:
\begin{itemize}
    \item We propose a framework to systematically evaluate LPPMs.
    \item We propose an improved LPPM.
    \item We systematically evaluate a selection of LPPMs.
    \item We show the dependence of mechanisms on the scenario.
\end{itemize}

%% file: 02.background.tex
\section{Background}\label{background}

Location privacy is an important field of research. For years mechanisms have
been proposed to ensure privacy in continuous and sparse scenarios. In sparse
scenarios geolocation points are supposed to be independent from each other,
whereas in continuous scenarios there is a dependence between those points,
\eg{} they belong to a trajectory.

There are two main approaches in location privacy for the user and the
adversary. First there is the anonymization approach: for the user it means
anonymizing their data, for the adversary it means getting back to the identity
of the owner of the data. Then there is the obfuscation approach: for the user
it means obfuscating their data, for the adversary it means getting back to the
original data. In this paper, we focus on obfuscation and de-obfuscation
mechanisms, but for the sake of completeness we also introduce anonymization and
de-anonymization in this section.

\subsection{Location Privacy Preserving Mechanisms (LPPMs)}

\begin{table}[!ht]
\centering
\begin{tabular}{|l|l|l|}
\hline
\textbf{Mechanism}           & \textbf{Type} & \textbf{Density} \\ \hline
k-anonymity cloaking regions & Anonymization & Sparse         \\ \hline
k-anonymity dummy locations  & Anonymization & Sparse         \\ \hline
Promesse                     & Obfuscation   & Continuous       \\ \hline
PL geoind                    & Obfuscation   & Sparse         \\ \hline
Adaptive geoind              & Obfuscation   & Continuous       \\ \hline
Clustering geoind            & Obfuscation   & Continuous       \\ \hline
Memory clustering geoind     & Obfuscation   & Continuous       \\ \hline
\end{tabular}
\caption{Mechanisms overview}
\label{tab:mechanisms}
\end{table}

LPPMs are mechanisms that transform geolocation data such that it becomes more
private while losing as little utility as possible. There are different goals
for a LPPM: anonymization or obfuscation. Anonymization mechanisms do not change
the location of the user, but they anonymize it by making sure the adversary
cannot map back a user's identity to their locations. Obfuscation mechanisms
change the location of the user to make sure the adversary cannot map back a
user to their precise location.

\subsubsection{$k$-anonymity}

The principle of $k$-anonymity is to group the data of a user with that of at
least $k-1$ other ones with the same
quasi-identifier~\cite{samarati_protecting_1998}. A quasi-identifier is a set of
attributes chosen such that it corresponds to at least $k$ users. These
attributes are accessible to the adversary. A dataset is then said to be
$k$-anonymous if every record in the dataset is indistinguishable from at least
$k-1$ other records. $k$-anonymity seeks to ensure privacy by providing
anonymity: the record of a user is grouped with similar records and they are all
labeled by the quasi-identifier common to this group.

This principle was adapted to geolocation data to ensure its privacy by Gruteser
and Grunwald~\cite{gruteser_anonymous_2003}: a user is said to be $k$-anonymous
if their location information is indistinguishable from that of at least $k-1$
other users. $k$-anonymity is the most popular anonymization mechanism for
location privacy. There are two approaches to apply $k$-anonymity to geolocation
data: cloaking regions and dummy locations techniques. With the cloaking regions
approach, location information is given as ranges of coordinates and timestamps
such that the location information of at least $k$ users fit into those
ranges~\cite{gruteser_anonymous_2003}. The range of coordinates correspond to
the quasi-identifier of the group. With the dummy locations
approach~\cite{kido_anonymous_2005}, the user reports $k-1$ dummy locations
alongside their real location. No additional information is given such that the
user's location is indistinguishable from the dummy locations. Implementing
$k$-anonymity does not require a lot of computational power. In cloaking
$k$-anonymity, the mechanism just has to create zones that contain at least $k$
locations. In dummy locations $k$-anonymity, the mechanism draws the dummy
points at random in the neighborhood of the previous
dummies~\cite{kido_anonymous_2005}. General $k$-anonymity (not specific to
location $k$-anonymity) is susceptible to homogeneity and background knowledge
attacks~\cite{machanavajjhala_l-diversity_2006}. Those attacks can be adapted
for location data: location homogeneity attack, map
matching~\cite{lamarca_inference_2007, wernke_classification_2014}. The
principle of $k$-anonymity was extended to notions such as
$l$-diversity~\cite{machanavajjhala_l-diversity_2006} or
$t$-closeness~\cite{li_t-closeness_2007}. These notions build upon $k$-anonymity
and add conditions to further ensure the privacy of the dataset. But
$k$-anonymity and those notions implicitly make assumptions on the adversary's
side information. For example in the dummy locations approach, the adversary is
assumed to not have any side knowledge allowing them to rule out the dummy
points~\cite{ghinita_preventing_2009}. As a result, it became more interesting
to look at approaches that abstract from the adversary's side information such
as differential privacy.

\subsubsection{Promesse: speed smoothing mechanism}

Primault et al. proposed \textit{Promesse}~\cite{primault_time_2015} a speed
smoothing mechanism to obfuscate user's locations in continuous scenarios. The
idea is to hide the user's Points Of Interests (POIs). POIs are determined by
grouping together points that are either close in space or close in time. To
this end, they not only obfuscate the location data but also the timestamps.
Their algorithm cannot be used to obfuscate data on the fly as it needs the full
mobility trace as input. From that mobility trace, they interpolate locations at
constant intervals of distance (and set the timestamps accordingly to smooth
speed overall) and blur the endpoints of the trace. The result is a location
trace that follows the real trajectory but where each point is situated at a
constant distance and time from the preceding and next ones.

\subsubsection{Notion of geo-indistinguishability}

Andres et al. introduced the notion of
geo-indistinguishability~\cite{andres_geo-indistinguishability_2013} as the
adaptation of the notion of differential
privacy~\cite{hutchison_differential_2006} to geolocation data. Differential
privacy is a statistic notion for privacy. A dataset is said to be
differentially private if modifying a subset of the records has a negligible
effect on its utility. The original dataset and the modified one are
indistinguishable.

A location privacy preserving mechanism $K$ is defined as a probabilistic
function assigning to each location $x$ a probability distribution ($K(x)$). In
order to apply differential privacy, any change in a location (from $x$ to $x'$)
should have a negligible effect on the published output (from $K(x)$ to
$K(x')$). Therefore, a LPPM $K$ satisfies $\epsilon$-geo-indistinguishability if
and only if for all $x$, $x'$~\cite{andres_geo-indistinguishability_2013}:

$$d_p\left(K(x),K(x')\right)\leq\epsilon d(x,x')$$

where $\epsilon$ is the geo-indistinguishability parameter defined as $\epsilon
= \ell/r$. Enjoying $\epsilon$-geo-indistinguishability means enjoying
$\ell$-privacy within a radius $r$. $\ell$ is the privacy level and $r$ is the
radius of the circle where the obfuscated point will be drawn.
$d\left(\cdot,\cdot\right)$ is the Euclidean distance and
$d_p\left(K(x),K(x')\right)$ is the distance between the corresponding
probability distributions.

\subsubsection{Planar Laplace geo-indistinguishability}

Andres et al.~\cite{andres_geo-indistinguishability_2013} proposed a first
geo-indistinguishable mechanism: the planar Laplace mechanism. It uses random
noise  drawn from a Laplace distribution to obfuscate all the geolocation
points. This mechanism apply planar Laplace noise to each point of the dataset
by adding $(r,\theta)$ noise to each point (in polar coordinates). For each
point, $\theta$ is drawn uniformly in $[0,2\pi)$ and $r$ is set to $r =
C_{\epsilon}^{-1}(p)$ where $p$ is drawn uniformly in $[0,1)$.

$$C_{\epsilon}^{-1}(p) =
-\frac{1}{\epsilon}\left(W_{-1}\left(\frac{p-1}{\epsilon}\right)+1\right)$$
where $W_{-1}$ is the $-1$ branch of the Lambert $W$ function. This mechanism is
an obfuscation one. Instead of anonymizing the data, this approach aims to
obfuscate the data so that the adversary cannot get precise information on the
user. It is considered as the state of the art for geolocation privacy by
obfuscation. Using the notion of geo-indistinguishability allows abstracting
from the adversary side knowledge. It is very efficient in sparse scenarios. But
in continuous scenarios, when geolocations are somehow correlated, it decreases
the level of privacy of the planar Laplace mechanism.

\subsubsection{Adaptive geo-indistinguishability}

The planar Laplace mechanism does not address the fact that in continuous
scenarios, points in the trajectory are correlated which will decrease the
overall privacy of the obfuscated trajectory. Al-Dhubhani and Cazalas addressed
this problem by proposing adaptive
geo-indistinguishability~\cite{al-dhubhani_adaptive_2018}. This mechanism uses
planar Laplace noise and constantly adapts the noise applied to the points to
increase the estimation error of the adversary while ensuring a minimum utility
to the user. To do that, the mechanism keeps the estimation error inside a
target range. In the following formula, $x$ is the real location, $\hat{x}$ is
the predicted obfuscated location obtained from a linear regression on the most
recent obfuscated points.

\[
\epsilon = \begin{cases}
\alpha\times\epsilon & \text{if } d(x,\hat{x})<\Delta_1\\
\epsilon & \text{if } \Delta_1\leq d(x,\hat{x})<\Delta_2\\
\beta\times\epsilon & \text{if } d(x,\hat{x})\geq\Delta_2
\end{cases}
\]
where $\Delta_1$ and $\Delta_2$ are bounds on the acceptable values for the
estimation error, $\alpha < 1$ and $\beta > 1$. The idea is to constantly
monitor the error level of the adversary by estimating what is their next
probable guess. If the estimation error of the adversary is too low, the
geo-indistinguishability parameter is reduced to add more noise and increase the
error level. If the estimation error is too high, the geo-indistinguishability
parameter is increased to reduce the error level. Although the added noise is
dependent on the estimation error, there is still a correlation between the
obfuscated points. It also does not deal well with obfuscating points of
interest.

\subsubsection{Clustering geo-indistinguishability}

The mechanism of clustering
geo-indistinguishability~\cite{cunha_clustering_2019} was developed for both
continuous and sparse scenarios. The idea is to create a report for a continuous
scenario that is the same than for a sparse scenario on the same trajectory. The
mechanism works as follows: it creates a cluster by obfuscating the location of
the user using planar Laplace noise and defining a cluster radius around the
real location of the user. Inside that cluster it will always report the same
obfuscated location. When out of the cluster, it will create a new one centered
around the new real location of the user and forget the old one.

By clustering the points, this mechanism reduces the diversity of information an
adversary could get, while the cluster radius ensures a minimum utility.
However, when the user does frequent and similar movements (such as commuting
every day), it is possible to average the points to get more precise locations
of where the user goes frequently (workplace, home) because the algorithm does
not keep in memory the previous clusters and will generate a new point each
time.

\subsection{Location Privacy Attacks}

The two main attacks on users' location privacy are tracking and identification
attacks~\cite{shokri_unied_2010}. Whereas identification attacks aim at
de-anonymizing location traces, tracking attacks aim at de-obfuscating location
traces. In this section we mainly focus on de-obfuscation attacks.
De-obfuscation mechanisms are mechanisms used by an adversary that take
obfuscated geolocation data and try to go back as close as possible to the
original location data.

\subsubsection{Multiple position attacks}
This class of attacks tracks the position updates of a user to correlate them
and find a more precise location of the user~\cite{wernke_classification_2014}.
Such attacks particularly targets $k$-anonymity. The shrink region attack and
region intersection attack~\cite{talukder_preventing_2010} calculate the
intersection of several imprecise locations of the user to decrease their
privacy and get a more precise estimate of the real position of the user.
Another attack, the maximum movement boundary
attack~\cite{ghinita_preventing_2009} reduces the region where the user could
be, by removing areas that could not be reachable given the maximum user's
speed.

\subsubsection{Probability distribution attack}
The attacker uses locations reports to create a probability distribution of the
user position. If the probability is not uniformly distributed, the attacker can
identify areas where the user could be. Shokri et
al.~\cite{hutchison_quantifying_2011} propose the Bayesian inference attack for
Hidden Markov Processes targeting sparse location exposure.

\subsubsection{Map matching}
Map matching algorithms aim at linking imprecise geolocation points to the most
coherent position on a map. It is very useful for GPS navigation systems. In the
case of de-obfuscation attacks, those algorithms allow to map back obfuscated
points to a coherent position on a map with the assumption that it will get the
points closer to their original position. A simple approach is to just map each
obfuscated point to the closest coherent point on the map. Newson and
Krumm~\cite{newson_hidden_2009} used Hidden Markov Model (HMM) to do the mapping
in order to take into account the coherence between the successive points in the
same trajectory.

Jagadeesh and Srikanthan~\cite{jagadeesh_online_2017} proposed an updated
implementation of a map matching algorithm. Considering a set of observations,
the HMM consists of all states in the surrounding area of each observation. In
practice the states are the road segments lying within a fixed range around each
observation. For each state, the emission probability is the probability that a
given observation is the point reported for that state. The transition
probability between two states is the probability of going from the first state
to the second one given the optimal real path and the temporal condition. With
the Viterbi algorithm~\cite{jagadeesh_online_2017}, it is then possible to
compute the most likely sequence of states in the HMM using emission and
transition probabilities.

\subsubsection{POI extraction}
Primault et al.~\cite{primault_differentially_2014} used point of interest
(POI) extraction on the obfuscated data to find the POI of the original data.
They found that this attack could identify the original POIs with a good success
rate. Given a maximum distance and a minimum time period, they group consecutive
points that are close to each other (less than the maximum distance) and only
keep the groups that span over more than the minimum time period.

\subsubsection{Linear regression}
In adaptive geo-indistinguishability~\cite{al-dhubhani_adaptive_2018}, the
estimation of the guess of the adversary is done by making a linear regression
on the last few obfuscated points according to their timestamps.

\subsubsection{Sliding average}
In a continuous scenario, using a sliding average over the few last and next
points can approximate the original trajectory. However, in some cases such as a
sharp turn, it can increase the level of noise instead of reducing it.

\subsection{Metrics}

The following metrics describe the level of privacy that the obfuscated dataset
provides compared to the original one~\cite{wagner_technical_2019}.

\subsubsection{Average Error}
Computing the average Euclidean distance between the original points and their
obfuscated version gives the error in utility for the user. Similarly, the
average distance between original points and their de-obfuscated version gives
the error in the estimation of the adversary. In the case of $k$-anonymity when
using this metric, distance is estimated with the closest point of the region
(or the border of the region)~\cite{ghinita_preventing_2009}.

\subsubsection{F-score}
Jagadeesh and Srikanthan~\cite{jagadeesh_online_2017} introduced this metric
which looks at how similar the trajectories are. It computes the length where
the de-obfuscated path overlaps the original one ($L_{correct}$) and compares it
to the total length of the original ($L_{truth}$) and de-obfuscated paths
($L_{matched}$).

$$precision = \frac{L_{correct}}{L_{matched}}$$
$$recall = \frac{L_{correct}}{L_{truth}}$$
$$F\text{-}score = 2\times\frac{precision\times recall}{precision + recall}$$

A high score means that the de-obfuscated path is similar to the original one.

\subsubsection{($\alpha$, $\delta$)-usefulness}
This measures the usefulness of a privacy mechanism. Andres et
al.~\cite{andres_geo-indistinguishability_2013} defined it as follows: a
mechanism $K$ is $(\alpha, \delta)$\textit{-useful} if for every location $x$,
the reported location $z = K(x)$ satisfies $d(x, z)\leq\alpha$ with probability
of at least $\delta$.

\subsubsection{POI related metrics}
Primault et al.~\cite{primault_differentially_2014} used POI extraction as a
de-obfuscation mechanism. They then mapped each of the obfuscated POIs to the
closest original POI. The metrics they used were the distance between the
obfuscated POIs and the original POI they are linked to, and the proportion of
original POIs that have an obfuscated POI mapped to it.

\subsection{Related Work}

Configuring LPPMs could be difficult and papers have proposed frameworks to make
configuration easier. Shokri et al.~\cite{shokri_unied_2010} propose a framework
to provide structure for classifying and organizing components and concepts of
location privacy. The goal of this framework is to help better understand the
field of research, identify problems, design new LPPMs and compare existing
LPPMs from a conception point of view. Their framework does not allow comparing
existing LPPMs on their real performance by applying them on real data. This
framework has a more abstract approach and models the components to compare them
from an architecture point of view. Shokri et
al.~\cite{hutchison_quantifying_2011} leverage on this framework to provide
higher location privacy for the users in sparse location exposure. Primault et
al.~\cite{primault_adaptive_2016} propose a framework for evaluating and
dynamically configuring LPPMs. The dynamic configuration is automatic and faster
than a manual one. The user sets some privacy and utility objectives that the
LPPM should satisfy, and the framework optimizes the parameters of the LPPM.
Cerf et al.~\cite{cerf_toward_2016} propose another framework to help the
fine-tuning of LPPMs according to some privacy and utility expectations. Both
frameworks are oriented toward the configuration of the LPPMs, they do not allow
to easily and systematically compare all mechanisms.

All these works although of high quality in their own right, performed their
evaluation in isolation and targeted towards specific scenarios to demonstrate
the utility of their mechanisms. They did not evaluate it over different
scenarios, different attack vectors nor did they look at it with respect to
datasets.

%% file: 03.methodology.tex
\section{Methodology}\label{methodology}

\begin{figure*}[!ht]
    \centering
    \includegraphics[width=.7\linewidth]{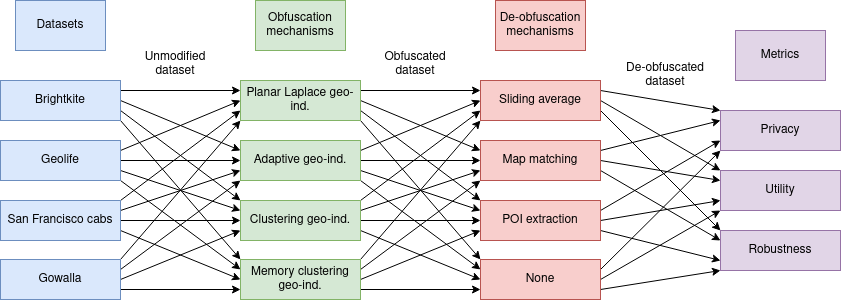}
    \caption{Framework diagram}
    \label{fig:framework}
\end{figure*}

\subsection{Definitions}

By looking into research papers about location privacy mechanisms, it appears
that some components are common when evaluating a LPPM and can be gathered in
four categories: datasets, obfuscation mechanisms, de-obfuscation mechanisms,
and metrics.

The datasets are sets of geolocation records with timestamp and identifier of
the corresponding user. Datasets can have different natures according to their
attributes: they can be sparse or dense from a spatial or temporal point of
view. In the rest of this paper, we call scenario a dataset with all its
attributes.

The obfuscation mechanisms are algorithms that apply on raw geolocation data and
add some noise to each point to reduce the precision of what an adversary could
get. The obfuscated data is what is being sent by the user to the LBS. It
defines what utility they will get and what the adversary will use to try
to locate them.

The de-obfuscation mechanisms are algorithms used by an adversary to try
reducing the noise added during the obfuscation mechanism. The goal is to use
the obfuscated data to get a more precise trajectory of the user or a more
precise idea of some locations of the user (\eg{} points of interest).

In this threat model, the user shares the obfuscated data with the LBS and the
adversary is considered to have access to this obfuscated data. The metrics
evaluate the privacy or utility levels of the mechanisms. They compare a set of
data (obfuscated or de-obfuscated) to the original data to compute the metric.

\subsection{Approach}

When introducing a new location privacy mechanism, we are interested in its
performance and its capability to ensure some privacy guarantees. Moreover, in
this work we are interested in the sensitivity of privacy mechanisms to the
scenario on which they are used. In order to study this, we need an evaluation
framework that let us evaluate mechanisms in different scenarios.

We want to evaluate the performance of the privacy mechanisms according to the
three dimensions commonly identified in the literature: privacy, robustness, and
utility. Privacy is measured after applying an obfuscation mechanism: it
represents how much information the obfuscation mechanism was able to suppress.
A high level of privacy means that the obfuscated data does not contain much
information. Robustness is the level of privacy we get after applying a
de-obfuscation mechanism: it corresponds to the level of privacy a mechanism is
able to maintain in the presence of an adversary. Utility is the amount of
information contained in the data: it corresponds to the degree to which the
data is useful. In this section, we present the evaluation framework we
developed in order to systematically evaluate the performance of LPPMs
(\autoref{fig:framework}). The framework consists of the different layers we
identified earlier: dataset, obfuscation mechanism, de-obfuscation mechanism,
and metric. An evaluation using this framework consists in choosing one element
from each layer and linking them together to run the evaluation. Using a common
interface between layers allows to link the different elements of one layer to
the ones of the next layer. By running each unique combination, this
architecture allows to systematically evaluate an exhaustive combination of
datasets, obfuscation mechanisms, de-obfuscation mechanisms, and metrics.

In this paper, we focus on a selection of elements for each of the layers. The
first layer is the dataset layer. The datasets allow us to test the mechanisms
on realistic data. For this paper, we choose two continuous datasets and two
sparse ones. The datasets define the scenario on which the mechanisms will be
evaluated. The datasets used in this paper are listed in
\autoref{tab:framework_components}. The obfuscation mechanism layer corresponds
to the LPPMs being tested. A LPPM takes as input a dataset and returns another
dataset more private than the original one and with minimum loss on utility. For
this paper, we focus on different geo-indistinguishability mechanisms as they
are considered the state of the art for location privacy especially for
continuous scenarios. The obfuscation mechanisms used in this work are listed in
\autoref{tab:framework_components}. The de-obfuscation mechanism layer
corresponds to the attacks that aim at reducing the obfuscation in order to get
back as close as possible to the original data. In this paper, we focus on three
different de-obfuscation mechanisms listed in
\autoref{tab:framework_components}. The metric layer corresponds to the metrics
evaluating the obfuscation or de-obfuscation mechanism to assess their privacy
or utility level. As explained earlier, we are interested in measuring the
privacy, robustness and utility levels of the chosen LPPMs. In this paper, we
focus on the metrics listed in \autoref{tab:framework_components}. The average
error metric gives us the privacy level of the data when applied on obfuscated
data, and the robustness level when applied on de-obfuscated data. POI recall
gives us the utility of the data it is applied on. ($\alpha$,
$\delta$)-usefulness can be used to measure both the utility and privacy levels.
Robustness corresponds to the level of privacy of the obfuscation mechanism
given by these privacy metrics when an attacker is considered. The POI metric
can only be used after the POI de-obfuscation mechanism. We choose to consider
POI extraction both as a de-obfuscation mechanism and as the first step of the
POI metric. Therefore, we can compute the POI metric after every mechanism by
first running POI extraction on the data.

\begin{table*}[!ht]
\centering
\resizebox{\linewidth}{!}{%
\begin{tabular}{|l|l|l|l|}
\hline
\textbf{Datasets}           & \textbf{Obfuscation mechanisms}
& \textbf{De-obfuscation mechanisms} & \textbf{Metrics}
\\ \hline
Brightkite         & Planar Laplace geo-indistinguishability    & Sliding
average           & Average error (Privacy)                  \\
Geolife            & Adaptive geo-indistinguishability          & POI extraction
& POI recall (Utility)                \\
Gowalla            & Clustering geo-indistinguishability        & Map matching
& ($\alpha$, $\delta$)-usefulness (Utility and Privacy) \\
San Francisco Cabs & Memory clustering geo-indistinguishability &
&                                 \\ \hline
\end{tabular}%
}
\caption{Frameworks elements}
\label{tab:framework_components}
\end{table*}

\subsection{Implementation}

The framework is implemented using Python. The dataset class is mainly an array
containing the data with some custom methods to make the processing easier. Each
dataset is pre-processed to make sure the format is consistent and to apply some
filtering (\eg{} removing trajectories too short, or points too far from the
center point of the study):
\begin{itemize}
    \item Geolife: the trajectories of the days when there are less than 480
    points are dropped
    \item San Francisco: the points when the cab is empty are dropped
    \item Brightkite: the users with less than 25 points are dropped
    \item Gowalla: the users with less than 50 points are dropped
\end{itemize}

Obfuscation and de-obfuscation mechanisms are functions applying on the dataset
class. We implemented them following the algorithms given in the corresponding
paper if any.

For map matching the implementation uses parts of modules introduced by
Boeing~\cite{boeing_osmnx_2017} and Mermet and Dujardin~\cite{mermet_etat_2020}.
We implement a map matching algorithm that maps each point to the closest node
in the OpenStreetMap network. Given that Brightkite and Gowalla points are all
over the world, running our map matching algorithm on these datasets would take
too much time (retrieving and going through the OpenStreetMap network for the
entire world). We decide not to run map matching on these datasets.

When extracting the points of interest, the algorithm groups points that are
within 250 m from each other. Because the smallest threshold used for the
spatial sub-sampling is 500 m, we decide not to run POI extraction and POI
metric on the spatially sub-sampled datasets as it will not be able to extract
any POI.

Metrics take as input both the original dataset and the one being evaluated in
order to compute measurements on the privacy or utility levels. We implement
them using the definitions given in published papers. Because of the data we get
after applying POI extraction, we cannot compute the $(\alpha,
\delta)$-usefulness after this de-obfuscation mechanism.

\subsection{Memory Clustering Geo-indistinguishability}

The framework can be extended by anyone. We use this capability to implement our
own mechanism and include it in our study to compare it to the other mechanisms.

We propose a new location privacy mechanism. This mechanism builds upon the
principle of the clustering geo-indistinguishable mechanism proposed by Cunha et
al.~\cite{cunha_clustering_2019}, but we add a memory to the algorithm (see
\autoref{alg:mem_geoind}). In clustering geo-indistinguishability, the current
point that we want to obfuscate is compared to the previous one to check if it
is in the same cluster. By adding memory, we can check that the current point we
want to obfuscate is not part of any of the previously created clusters.

Using simple clustering geo-indistinguishability, if we consider a user doing a
recurrent trip involving at least two recurrent destinations (for example
commuting between home and work every day), then because the comparison only
checks the last point, each time the user reaches one of the destination, a new
point is obfuscated and reported. Thus using their location records over a few
days could allow narrowing down their points of interests.

In the same scenario, using clustering geo-indistinguishability with memory,
each time the user reaches one of its destinations, because the previous reports
are kept in memory, no new point will be reported.

\begin{algorithm}[!ht]
\SetKwProg{Fn}{Function}{ :}{end} \SetKwFunction{MC}{MemoryClustering}
\SetKwFunction{PL}{PlanarLaplace} \SetKw{Ret}{return} \Fn{\MC($real\_mem =
\{x_1,\dots, x_n\}$, $obf\_mem = \{z_1,\dots, z_n\}$, $x_c$, $\epsilon$, $r$)}{
\eIf{$real\_mem = \emptyset$}{ $z_c \gets$ \PL{$x_c, \epsilon$}\\
        $real\_mem \gets \{x_c\}$\\
        $obf\_mem \gets \{z_c\}$ }{ $idx = argmin_{i \in
    \{1,\dots,n\}}\left(d(x_c, x_i)\right)$\\
        \eIf{$d(x_c, x_{idx}) \leq r$}{ $z_c \gets z_{idx}$ }{ $z_c \gets$
            \PL{$x_c, \epsilon$}\\
            $real\_mem \gets real\_mem \cup \{x_c\}$\\
            $obf\_mem \gets obf\_mem \cup \{z_c\}$ } } \Ret{$z_c$} }
        \caption{Memory clustering geo-indistinguishability
        algorithm}\label{alg:mem_geoind}
\end{algorithm}

%% file: 04.evaluation.tex
\section{Evaluation}\label{evaluation}

We focus our evaluation around the three aspects of the performance of a LPPM:
privacy, robustness, and utility. Accordingly, we ask the following evaluation
questions:
\begin{itemize}
    \item Is there a most private obfuscation mechanism?
    \item Is there a most efficient obfuscation mechanism?
    \item Is there a most robust obfuscation mechanism?
\end{itemize}

\paragraph*{\textbf{Roadmap}}

To answer these questions, we run 4 different experiments (see
\autoref{tab:roadmap}) to evaluate the privacy and utility of these obfuscation
mechanisms on the original (Experiment 1) and sub-sampled (Experiment 2)
datasets. We then study the robustness of these mechanisms on the original
(Experiment 3) and sub-sampled (Experiment 4) datasets. The insights and main
takeaways gained from these experiments are summarized at the end of each
corresponding section.

\begin{table}[!ht]
\resizebox{\linewidth}{!}{%
\begin{tabular}{c|cc|}
\cline{2-3}  & \multicolumn{2}{c|}{\textbf{Metrics considered}} \\

 \cline{2-3}  & \multicolumn{1}{c|}{\textbf{Privacy and Utility}} &
 \textbf{Robustness} \\
 \hline

 \multicolumn{1}{|c|}{\textbf{Original Datasets}} &
\multicolumn{1}{c|}{Experiment 1 (\autoref{exp1})} & Experiment 3
(\autoref{exp3}) \\

\cline{1-3}  \multicolumn{1}{|c|}{\textbf{Sub-sampled Datasets}} &
\multicolumn{1}{c|}{Experiment 2 (\autoref{exp2})} & Experiment 4
(\autoref{exp4})\\ \hline
\end{tabular}
}
\caption{Roadmap of the evaluation experiments}
\label{tab:roadmap}
\end{table}

\begin{table*}[!ht]
    \centering
    \resizebox{\linewidth}{!}{%
    \begin{tabular}{|l|l|l|l|l|l|l|l|l|}
    \hline
    \textbf{Dataset}          & \begin{tabular}[c]{@{}l@{}}\textbf{Number of}\\
    \textbf{users}\end{tabular} &
    \begin{tabular}[c]{@{}l@{}}\textbf{Number}\\ \textbf{of
    points/user}\end{tabular} &
    \begin{tabular}[c]{@{}l@{}}\textbf{Distance between two}\\
    \textbf{consecutive points ($m$)}\end{tabular} &
    \begin{tabular}[c]{@{}l@{}}\textbf{Frequency}\\ \textbf{($Hz$)}\end{tabular}
    &
    \begin{tabular}[c]{@{}l@{}}\textbf{Velocity}\\ \textbf{($m/s$)}\end{tabular}
    &
    \begin{tabular}[c]{@{}l@{}}\textbf{Time window}\\
    \textbf{(days)}\end{tabular} &
    \begin{tabular}[c]{@{}l@{}}\textbf{Density}\\
    \textbf{(points/$km^2$)}\end{tabular} & \textbf{Area ($km^2$)} \\ \hline
    San Francisco Cab & 536
    & 9958                                                            & 654 &
    0.0183                                                     & 8.17 & 23
    & 13 & 780           \\ \hline
    Geolife           & 178
    & 37400                                                           & 15 &
    0.4404                                                     & 4.12 & 72
    & 130 & 404           \\ \hline
    Gowalla           & 196,591
    & 98                                                              & 35445 &
    0.0012                                                     & 2.69 & 143
    & 0.00034 & 324,000       \\ \hline
    Brightkite        & 58,228
    & 87                                                              & 57965 &
    0.0025                                                     & 4.98 & 338
    & 0.000057 & 1,050,000     \\ \hline
    \end{tabular}%
    }
    \caption{Median values of the attributes of the datasets}
    \label{tab:attributes}
\end{table*}

\subsection{Experimental Setup}
\subsubsection{Original datasets}

Having real location data is very useful for evaluating LPPMs. But collecting
location data could be challenging and expensive for different reasons such as
privacy even for research. For that reason, a few datasets have been made
publicly available and are commonly used for location data related research.

\paragraph*{\textbf{Geolife}} The Geolife dataset is a collection of
trajectories mainly in the Beijing area (China) collected by Microsoft Research
Asia~\cite{zheng2011geolife} between April 2007 and October 2011. The data was
collected over 178 users in a broad range of situations such as commute
trajectories, hiking or cycling.

\paragraph*{\textbf{San Francisco Cabs}} The San Francisco Cabs dataset is a
collection of trajectories of taxis in the San Francisco Bay Area (USA)
collected during one month in 2008. It is made publicly available through the
CRAWDAD project~\cite{epfl-mobility-20090224}. It contains mobility traces for
approximately 500 taxis.

\paragraph*{\textbf{Brightkite and Gowalla}} Brightkite and Gowalla were two
location-based social networking services. The datasets consist of the
geolocations of users of these social networks~\cite{cho_friendship_2011}. The
geolocation data are very sparse given that the locations were only recorded
when the users checked-in on mobile applications or websites. The locations are
spread all around the world. The Brightkite dataset was collected between April
2008 and October 2010 over 58,228 users; the Gowalla dataset was collected
between February 2009 and October 2010 over 196,591 users.

\paragraph*{\textbf{Attributes of the original datasets}}

We want to get a more precise idea of how the datasets differ. We identify
possible attributes for each dataset (\autoref{tab:attributes}):
\begin{itemize}
    \item Number of points per user: the San Francisco and Geolife datasets
    users report more points than the users of Gowalla and Brightkite datasets;
    \item Median distance between two consecutive points for each user: points
    in the Geolife dataset are 15 m apart, whereas points in the Brightkite
    dataset are 58 km apart;
    \item Median frequency for each user: a user of the Geolife dataset reports
    a point every 2 seconds, whereas a user of the Gowalla dataset reports a
    point every 14 minutes;
    \item Median velocity of each user: velocities are similar for all datasets;
    \item Time window of each user: the time windows of the San Francisco
    dataset are the shortest (23 days median); the time windows of the
    Brightkite dataset are the longest (338 days median);
    \item Density of each user (number of points per area): Brightkite and
    Gowalla are not dense compared to the San Francisco and Geolife datasets;
    \item Area covered by each user: the areas covered by users of the
    Brightkite and Gowalla datasets are very large.
\end{itemize}

From these attributes, we can identify two categories of datasets: the
continuous ones (San Francisco and Geolife) and the sparse ones (Gowalla and
Brightkite). The two attributes that are the most important for this distinction
are the frequency of report and the distance between two consecutive points.
They allow to characterize the sparsity of a dataset from a spatial and temporal
point of view.

\subsubsection{Sub-sampled datasets}
We decide to sub-sample the San Francisco Cabs dataset in order to evaluate the
mechanisms on datasets with specific attributes. We choose this dataset because
it is a continuous one, and we can easily sub-sample it. We decide to focus on
two attributes of the dataset: the frequency of report and the distance between
two consecutive points.

From the temporal point of view the sub-sampled datasets are a dataset where two
consecutive points are separated by at least: 1 min, 10 min, 30 min, and 1 h.
From the spatial point of view the sub-sampled datasets are  a dataset where two
consecutive points are separated by at least: 500 m, 1 km, 5 km, and 10 km.

\subsubsection{Parameters}

\paragraph*{\textbf{Use cases}}

We define three real-life use cases where a user could be using LBSs with
different utility level:
\begin{enumerate}
    \item \textbf{Vehicle for hire}: when a user request a ride to a vehicle for
    hire service (such as Uber or Lyft), the position of the user is used to
    match them with a nearby car and set the pickup point; we consider that the
    user does not mind walking to the pickup point if it is close to their real
    location; the acceptable level of noise for this use case is between 100~m
    and 500~m.
    \item Local businesses and restaurants: in this use case, the user want to
    retrieve some information about local businesses around them from services
    such as Google Maps or TripAdvisor; this type of requests often covers an
    area such as a neighborhood; the acceptable level of noise for this use case
    can go up to 1~km.
    \item Weather forecast and local news: for this type of services, the
    granularity is at the level of a city; the acceptable level of noise for
    this use case can go up to 10~km.
\end{enumerate}

\paragraph*{\textbf{$\epsilon$ values}}

We want to evaluate the mechanisms with parameters coherent with a real-world
use case. The first values to define are the values of $\epsilon$ for the
geo-indistinguishability mechanisms. We simulate the amount of noise (in meters)
added to a point using Planar Laplace noise for different values of $\epsilon$
(\autoref{tab:epsilons}). Given the use cases we defined, we want to use values
of $\epsilon$ that add noise within the range 100~m - 10~km . We choose to use
similar values to the ones used by Primault et
al.~\cite{primault_differentially_2014}. They define three values of $\epsilon$
for strong, medium, and weak privacy. For the first part of our evaluation, we
select the strong and weak values of $\epsilon$ and sub-sample a range of 10
values of $\epsilon$ in between those two values (\autoref{tab:my_eps}).

For the second part of the evaluation we only consider the three values of
$\epsilon$ introduced by Primault et al.:
\begin{itemize}
    \item Strong privacy: $\epsilon = 0.00139$~m$^{-1}$;
    \item Medium privacy: $\epsilon = 0.00358$~m$^{-1}$;
    \item Weak privacy: $\epsilon = 0.00693$~m$^{-1}$.
\end{itemize}

\begin{table}[!ht]
\centering
\begin{tabular}{|l|l|l|}
\hline
$\epsilon$ ($m^{-1}$) & Average noise ($m$) & Maximum noise ($m$) \\ \hline
0.00050               & 4004.1              & 29255.1             \\ \hline
0.00100               & 1999.3              & 14675.6             \\ \hline
0.00500               & 399.1               & 3069.9              \\ \hline
0.01000               & 200.7               & 1322.5              \\ \hline
0.05000               & 39.9                & 253.6               \\ \hline
0.10000               & 19.9                & 177.8               \\ \hline
0.50000               & 4.0                 & 27.4                \\ \hline
1.00000               & 2.0                 & 15.0                \\ \hline
5.00000               & 0.4                 & 2.9                 \\ \hline
\end{tabular}
\caption{Mapping between $\epsilon$ values and noise added}
\label{tab:epsilons}
\end{table}
\begin{table}[!ht]
\centering
\begin{tabular}{|l|l|l|}
\hline
$\epsilon$ ($m^{-1}$) & Average noise ($m$) & Maximum noise ($m$) \\ \hline
0.00139               & 1440.7              & 11532.3             \\ \hline
0.00200               & 1001.3              & 6557.5              \\ \hline
0.00262               & 763.6               & 5741.0              \\ \hline
0.00323               & 617.8               & 4119.4              \\ \hline
0.00385               & 519.9               & 4318.6              \\ \hline
0.00447               & 445.0               & 3103.2              \\ \hline
0.00508               & 391.9               & 2674.5              \\ \hline
0.00570               & 350.9               & 2683.7              \\ \hline
0.00632               & 316.8               & 2358.3              \\ \hline
0.00693               & 287.8               & 2300.0              \\ \hline
\end{tabular}
\caption{Values of $\epsilon$ used}
\label{tab:my_eps}
\end{table}

\paragraph*{\textbf{Other parameters}}

For the adaptive geo-indistinguishability mechanism, we use the values of the
original paper~\cite{al-dhubhani_adaptive_2018}:
\begin{itemize}
    \item $\Delta_1 = 693$~m and $\Delta_2 = 1948$ m
    \item $ws = 5$
    \item $\alpha = 0.1$ and $\beta = 5$
\end{itemize}

For clustering and memory clustering geo-indistinguishability mechanisms, we set
the clustering radius to 200 m. For the POI extraction mechanism, we group
together consecutive points such that the maximum diameter of the group is 250~m
and the user spent at least 1 h in that location.

\subsection{Experiment 1}
\label{exp1}

\textbf{Experiment 1} consists in evaluating the privacy and utility of the
considered obfuscation mechanisms on all original datasets.

\begin{figure}[!ht]
    \centering
    \includegraphics[width=.8\linewidth]{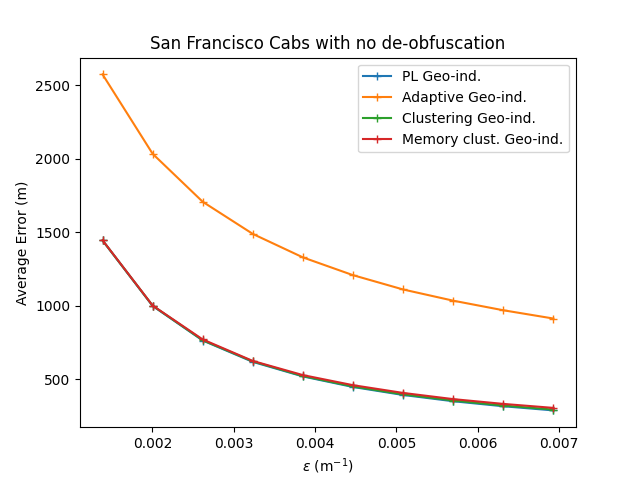}
    \caption{Average error for the San Francisco Cabs dataset (some lines are on top of each other)}
    \label{fig:sf_cabs_metric_avg_error}
\end{figure}

Looking at the average error we observe that the curves for PL, clustering, and
memory clustering geo-indistinguishability are very similar
(\autoref{fig:sf_cabs_metric_avg_error} shows the average error for the San
Francisco Cabs dataset). Because those mechanisms apply independently for each
point, what we observe is the average Planar Laplace noise for the specified
$\epsilon$ values. The adaptive geo-indistinguishability mechanism depends on
the distance and the period between the last points; that is why we can observe
that the curves differ from the other mechanisms for each dataset. The error we
get from this mechanism is mainly higher than the errors for the other
mechanisms: this mechanism is better for privacy in this scenario.

The POI metric gives us the recall of POIs from the original dataset that can be
mapped back from the obfuscated dataset POIs. Looking at
\autoref{fig:geolife_metric_POIMetric} we can observe that clustering and memory
clustering geo-indistinguishability lead to a high recall of the original POIs.
When using clustering or memory clustering geo-indistinguishability the points
the user reports are similar to the user's POIs in number and distribution.
Thus, it leads to a high recall. When using PL or adaptive
geo-indistinguishability, the points the user reports are spread along the
trajectory with some noise. It will be harder to compute the POIs from the
obfuscated data, thus leading to a lower recall. This is true for both
continuous and sparse datasets (\autoref{graphs_appendix}).

\begin{figure}[!ht]
    \centering
    \includegraphics[width=.8\linewidth]{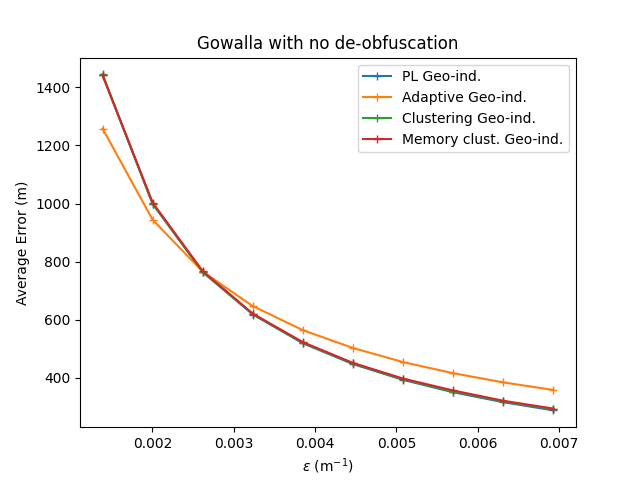}
    \caption{Average error for the Gowalla dataset (some lines are on top of each other)}
    \label{fig:gowalla_metric_avg_error}
\end{figure}

The ($\alpha$, $\delta$)-usefulness gives us the distribution of the distance of
the obfuscated points from the original point: for every distance $\alpha$
between 0 and 10~km , we compute the proportion $\delta$ of obfuscated points
that are within $\alpha$ of their original point. Using the use cases we
defined, we can observe the proportion of points that are useful for the user in
a given use case. We can observe (\autoref{graphs_appendix}) that PL,
clustering, and memory clustering geo-indistinguishability mechanisms behave
similarly for the different datasets. If we consider the local businesses use
case ($\alpha$ = 1~km ), for small values of $\epsilon$ (0.00139~m$^{-1}$) the
proportion of useful points is around 40\% and for high values of $\epsilon$
(0.00693~m$^{-1}$) the proportion of useful points is almost 100\%. The adaptive
geo-indistinguishability mechanism is dependent on the distance and the period
between the last points. We can observe this dependence, as it performs
differently for each dataset. On the Gowalla dataset the proportion of useful
points (for $\alpha$ = 1~km) is around 90\% for all $\epsilon$ values tested,
whereas on the Geolife dataset the proportion of useful points (for $\alpha$ =
1~km) is around 70\% for all $\epsilon$ values tested.

\begin{tcolorbox}[width=\linewidth, colback=white!95!black]
    \textbf{Takeaway:} The privacy performance of adaptive
    geo-indistinguishability is dependent on the scenario.
\end{tcolorbox}

\subsection{Experiment 2}
\label{exp2}

To gain a better understanding of the influence of the scenario on the privacy
and utility of the mechanisms, \textbf{Experiment 2} is run on the sub-sampled
datasets to study the temporal and spatial sampling effect.

\textbf{Experiment 2} consists in evaluating the privacy and utility of the
considered obfuscation mechanisms across different scenarios.

Looking at the average error for different values of sub-sampling, we observe
that as expected, only adaptive geo-indistinguishability is dependent on the
sub-sampling because the other mechanisms apply independently to each point
(\autoref{fig:spatial_sample_metric_avg_error} shows the average error for the
spatially sub-sampled datasets).

\begin{figure}[!ht]
    \centering
    \includegraphics[width=.8\linewidth]{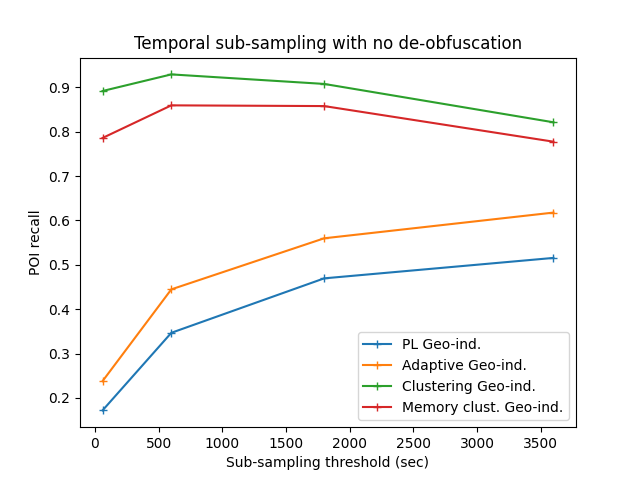}
    \caption{POI recall for the temporally sub-sampled datasets, for $\epsilon =
    0.00358$~m$^{-1}$}
    \label{fig:temp_sample_metric_POIMetric}
\end{figure}

The POI recall values of the sub-sampled datasets show that, as explained
previously, the clustering and memory clustering mechanisms lead to a high
recall of the original POIs (high utility).
\autoref{fig:temp_sample_metric_POIMetric} shows the POI recall for temporally
sub-sampled datasets for $\epsilon = 0.00358$~m$^{-1}$. We can observe that the
POI recall is higher for sparse datasets (20\% POI recall) for PL and adaptive
geo-indistinguishability than it is for continuous datasets (50\% POI recall).

If we consider the ($\alpha$, $\delta$)-usefulness graphs for the different
scenarios (\autoref{graphs_appendix}), we can observe that adaptive
geo-indistinguishability is the only mechanism to be dependent on the sampling
of the dataset. This is coherent with the observations made on the original
datasets.

\begin{tcolorbox}[width=\linewidth, colback=white!95!black]
    \textbf{Takeaway:} The privacy and utility performance of the mechanisms is
    dependent on the scenario. Clustering and memory clustering have the highest
    utility.
\end{tcolorbox}

\subsection{Experiment 3}
\label{exp3}

To assess the robustness of the obfuscation mechanisms, we carry out in
\textbf{Experiment 3} de-obfuscation attacks on the original datasets obfuscated
in different ways accordingly to the mechanism considered.

For the sliding average attack (\autoref{graphs_appendix}), we can observe that
for very sparse datasets (Brightkite and Gowalla) the average error is very high
(100s of km) compared to the denser datasets (San Francisco and Geolife). This
can be explained by the de-obfuscation mechanism used. We ran the sliding
average mechanism by averaging points before and after the current one. Thus, in
sparse scenarios, this attack will greatly move the points, whereas in more
continuous scenarios the points will not move that much.

The POI recall after a sliding average attack is higher for the PL and adaptive
mechanisms (between 60 and 80\%) for continuous datasets compared to their
recall without any de-obfuscation. For sparse datasets, the POI recalls are
overall around 60\% for all mechanisms.

For the map matching attack, the average error is bigger (8-10~km) than for
sliding average attack (1-2~km ). One explanation could be that the algorithm
simply link the points to the nearest possible point in the OpenStreetMap
network. Therefore, the points are not necessarily getting closer to the
original points after the map matching.

For the POI extraction attack, the average error we observe is on the order of
1~km  for continuous datasets but on the order of 20-50~km for sparse datasets.
The POIs from the de-obfuscated data are systematically mapped to the closest
original POI, therefore for sparse datasets it tends to map them back to POIs
that are not as close as the POIs from a continuous dataset.

\begin{figure}[!ht]
    \centering
    \includegraphics[width=.8\linewidth]{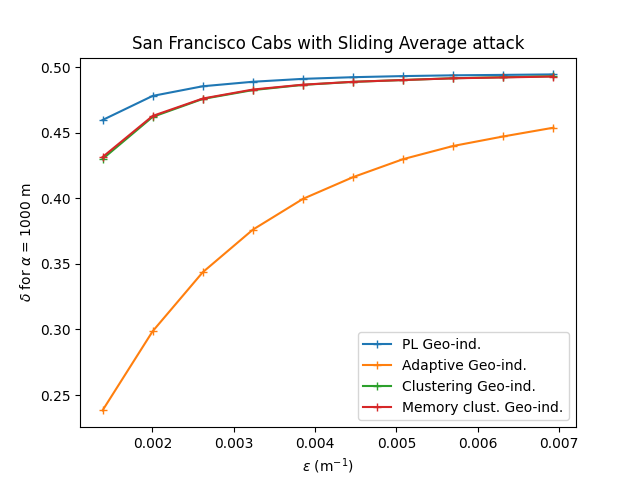}
    \caption{($\alpha$, $\delta$)-usefulness on San Francisco Cabs dataset (sliding average attack and $\alpha$ = 1000 m)}
    \label{fig:usefulness_1000_sf_cabs_de_obf_sliding_metric_alpha_delta}
\end{figure}

\begin{figure}[!ht]
    \centering
    \includegraphics[width=.8\linewidth]{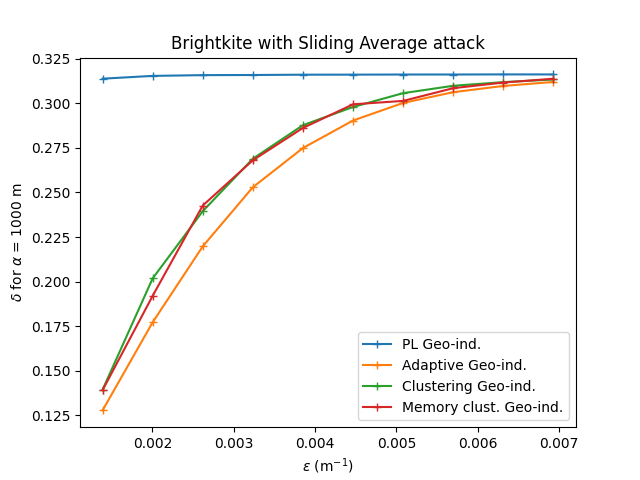}
    \caption{($\alpha$, $\delta$)-usefulness on Brightkite dataset (sliding average attack and $\alpha$ = 1000 m)}
    \label{fig:usefulness_1000_brightkite_de_obf_sliding_metric_alpha_delta}
\end{figure}

For the ($\alpha$, $\delta$)-usefulness results, we can see that overall the
proportion of useful points after a sliding average attack is higher for
continuous datasets
(\autoref{fig:usefulness_1000_sf_cabs_de_obf_sliding_metric_alpha_delta}) than
for sparse datasets
(\autoref{fig:usefulness_1000_brightkite_de_obf_sliding_metric_alpha_delta}). If
we consider the local businesses use case ($\alpha$ = 1~km), for all $\epsilon$
values tested, the proportion of useful points is higher for Geolife and San
Francisco Cabs datasets. We can even observe a difference between the Geolife
dataset (almost 100\% of useful points for high values of $\epsilon$) and the
San Francisco Cabs dataset (40-50\% of useful points for high values of
$\epsilon$).

\begin{tcolorbox}[width=\linewidth, colback=white!95!black]
    \textbf{Takeaway:} If we consider the robustness with the ($\alpha$,
    $\delta$)-usefulness metric and after a sliding average attack, adaptive is
    the most robust mechanism. This is not true for all other attacks.
\end{tcolorbox}

\subsection{Experiment 4}
\label{exp4}

Similarly to what we did earlier in Experiment 2, we now want to study the
influence of the scenario on the robustness of the mechanisms. Therefore, we run
de-obfuscation attacks on the sub-sampled datasets in \textbf{Experiment 4}.

The POI recall after a sliding average attack goes from 90\% for more continuous
datasets to 60\% for sparser datasets
(\autoref{fig:temp_sample_de_obf_sliding_metric_POIMetric}). This confirms the
observation on the original datasets: robustness is higher for sparse scenarios.

\begin{figure}[!ht]
    \centering
    \includegraphics[width=.8\linewidth]{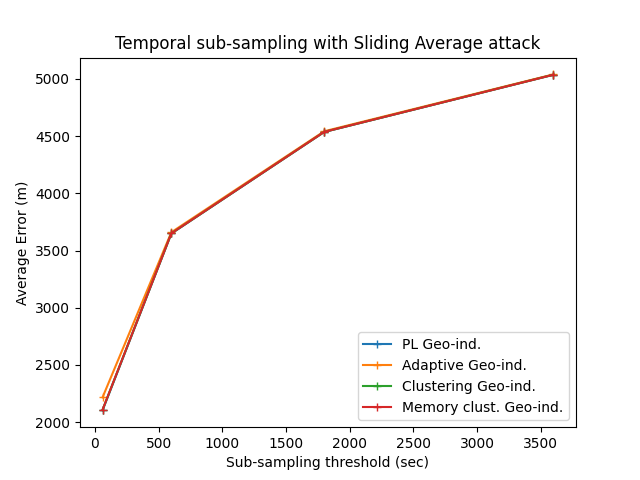}
    \caption{Average error for temporally sub-sampled datasets with sliding average attack (some lines are on top of each other)}
    \label{fig:temp_sample_de_obf_sliding_metric_avg_error}
\end{figure}

\begin{figure}[!ht]
    \centering
    \includegraphics[width=.8\linewidth]{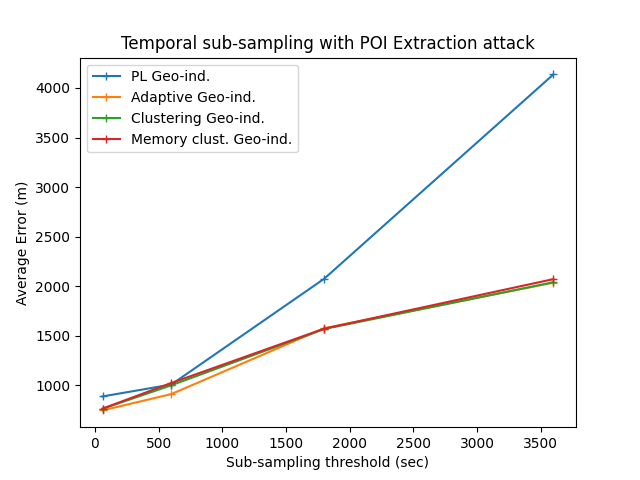}
    \caption{Average error for temporally sub-sampled datasets with POI extraction attack (some lines are on top of each other)}
    \label{fig:temp_sample_de_obf_poi_metric_avg_error}
\end{figure}

The average error after a POI extraction is higher for sparser datasets than it
is for more continuous ones
(\autoref{fig:temp_sample_de_obf_poi_metric_avg_error}). It confirms that for
sparse datasets, the POI extraction mechanism tends to match the POIs to
original POIs that are further than those for continuous datasets. In this
scenario, PL geo-indistinguishability is the most robust mechanism.

For spatially sub-sampled datasets, we observe that the mechanisms behave
similarly. This can be explained by the fact that the clustering radius for
clustering and memory clustering geo-indistinguishability is less than the
smallest threshold used to sub-sample the San Francisco Cabs dataset. Therefore,
these mechanisms cannot create clusters, and they behave just like PL
geo-indistinguishability.

The ($\alpha$, $\delta$)-usefulness results (\autoref{graphs_appendix}) for the
sub-sampled datasets confirm that more continuous datasets lead to higher
proportion of useful points after a sliding average attack. If we consider the
weather forecast use case ($\alpha$ = 10~km) denser datasets have a proportion
of almost 100\% of useful points, but for sparser datasets the proportion drops
to 30\% for spatial sub-sampling and 87\% for temporal sub-sampling.

\begin{tcolorbox}[width=\linewidth, colback=white!95!black]
    \textbf{Takeaway:} The robustness performance of the mechanisms is dependent
    on the scenario. There is no optimal mechanism.
\end{tcolorbox}

\subsection{Insights on the Mechanisms}
When looking at the results after applying an obfuscation mechanism, but without
any de-obfuscation mechanism, we observe that only adaptive
geo-indistinguishability is dependent on the scenario of the datasets. This is
explained by the way its algorithm works. Overall, the obfuscation mechanisms
provided comparable values for average error and ($\alpha$,
$\delta$)-usefulness. When looking at the POI recall values, we observe that
clustering and memory clustering geo-indistinguishability do not hide very well
the POIs compared to the other obfuscation mechanisms. In this regard, we can
say that clustering and memory clustering geo-indistinguishability are less
private than the other mechanisms. We also observe that PL and adaptive
geo-indistinguishability are less private for sparse datasets (50\% POI recall)
than they are for continuous datasets (20\% POI recall).

The results after running the de-obfuscation mechanisms show that overall the
error made by the adversary increases when the datasets become sparser for the
different attacks. For the POI recall, we observe that when a dataset becomes
sparser, the recall decreases after a sliding average attack.

Clustering and memory clustering are less robust against POI extraction attacks,
they tend to not hide as much POIs as the other mechanisms. PL
geo-indistinguishability is less robust than the other mechanisms against
sliding average attacks, while adaptive geo-indistinguishability is often the
most robust one in that situation. The privacy, utility, and robustness performances of the mechanisms are
dependent on the scenario.

%% file: 05.discussion.tex
\section{Discussion}\label{discussion}

\subsection{Limits}

We can identify some limits in our implementation of the framework. For example,
our implementation of the map matching algorithm cannot reasonably run on very
large datasets, and thus we only run it on the smallest datasets. Moreover, this
implementation is simple and could be improved by using Hidden Markov Model for
example. For performance reasons, our implementation of the POI extraction
mechanism could be optimized by computing the POIs in parallel.

\subsection{Future work}
The framework can be extended by construction, thus future work could add new
mechanisms: new LPPMs not necessarily based on geo-indistinguishability, and new
location privacy attacks.
Similarly, one could add new metrics to the framework (such as F1-score).

Future work could also be to make this framework more user-friendly. For
example, it could be an interactive visualization or website where you could
pick datasets with their scenario, mechanisms, and metrics and see the results.
It could be helpful for exploration and choice of which mechanism to use to help
manufacturers, engineers, etc., make a decision.

After the exploration of the mechanisms and their parameters, the next step is
the explainability: by decomposing each mechanism into smaller explainable
algorithm, we could for example understand the underlying reasons that make a
specific mechanism perform better than another one in a particular scenario.

This deeper understanding could then be leveraged to come up with new and
improved mechanisms for location privacy: an automatic shuffling of all
components (from the decomposition) could be used to craft maybe a better
algorithm.

%% file: 06.conclusion.tex
\section{Conclusion}\label{conclusion}

In this paper, we study the sensitivity of geo-indistinguishable obfuscation and
de-obfuscation mechanisms on the scenario on which they are used. To study this
sensitivity, we introduce an evaluation framework that allows us to
systematically evaluate LPPMs by considering an exhaustive combination of LPPMs,
attacks, and metrics. Using our framework we compare a selection of LPPMs
including an improved mechanism that we introduce. By evaluating over a variety
of scenarios, we find that the efficacy of the studied mechanisms vary according
to the scenario. For example, we find that the privacy of Planar Laplace
geo-indistinguishability is greatly reduced in a continuous scenario (20\%
points of interest recall) compared to a sparse scenario (45\% points of
interest recall). We show that the efficacy (privacy, utility, and robustness)
that obfuscation and de-obfuscation mechanisms provide is highly dependent on
the scenario on which they are used.

%% file: A.appendix-a.tex
\section{Graphs}\label{graphs_appendix}

\subsection{Brightkite}
\begin{figure}[h!]
    \centering
    \includegraphics[width=.8\linewidth]{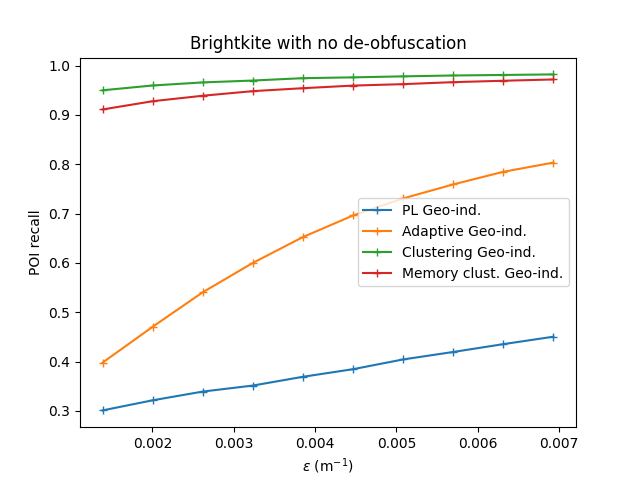}
    \caption{POI recall on Brightkite dataset}
    \label{fig:brightkite_metric_POIMetric}
\end{figure}
\begin{figure}[h!]
    \centering
    \includegraphics[width=.8\linewidth]{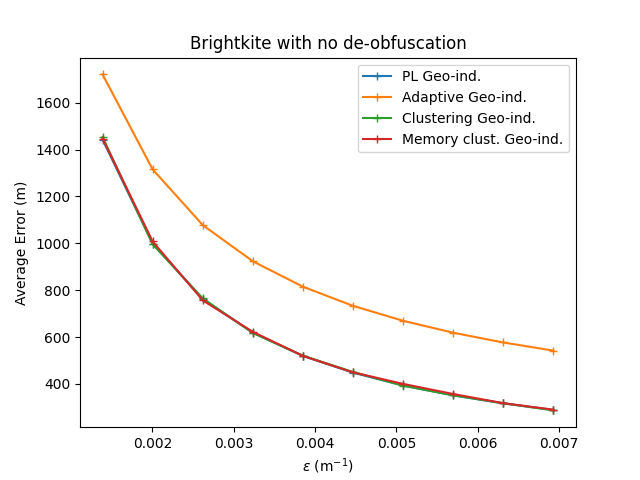}
    \caption{Average error on Brightkite dataset}
    \label{fig:brightkite_metric_avg_error}
\end{figure}

\begin{figure}[h!]
    \centering
    \includegraphics[width=.8\linewidth]{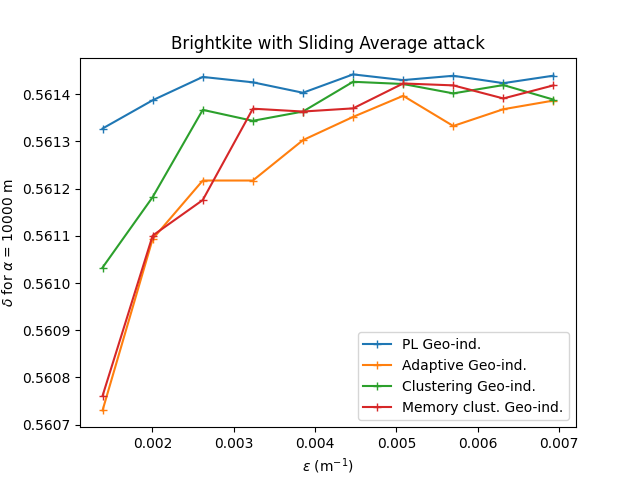}
    \caption{($\alpha$, $\delta$)-usefulness for $\alpha$ = 10~km on Brightkite dataset with sliding average attack}
    \label{fig:usefulness_10000_brightkite_de_obf_sliding_metric_alpha_delta}
\end{figure}

\FloatBarrier

\subsection{Geolife}
\begin{figure}[h!]
    \centering
    \includegraphics[width=.8\linewidth]{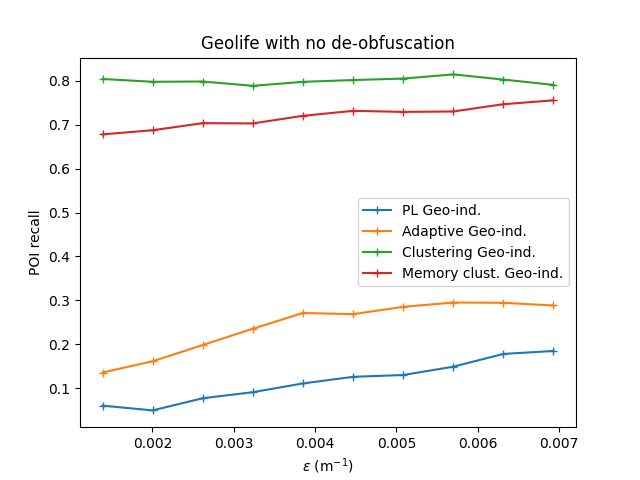}
    \caption{POI recall on Geolife dataset}
    \label{fig:geolife_metric_POIMetric}
\end{figure}

\begin{figure}[h!]
    \centering
    \includegraphics[width=.8\linewidth]{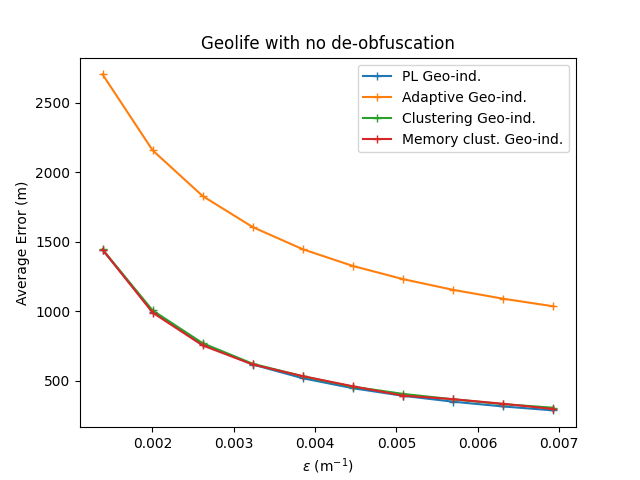}
    \caption{Average error on Geolife dataset}
    \label{fig:geolife_metric_avg_error}
\end{figure}
\begin{figure}[h!]
    \centering
    \includegraphics[width=.8\linewidth]{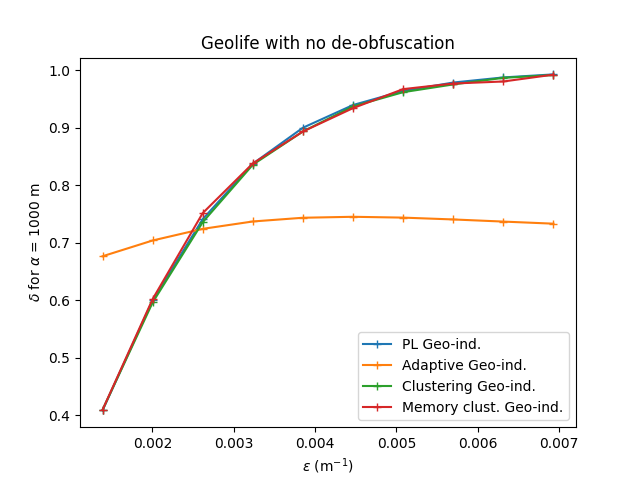}
    \caption{($\alpha$, $\delta$)-usefulness for $\alpha$ = 1~km on Geolife dataset}
    \label{fig:usefulness_1000_geolife_metric_alpha_delta}
\end{figure}
\begin{figure}[h!]
    \centering
    \includegraphics[width=.8\linewidth]{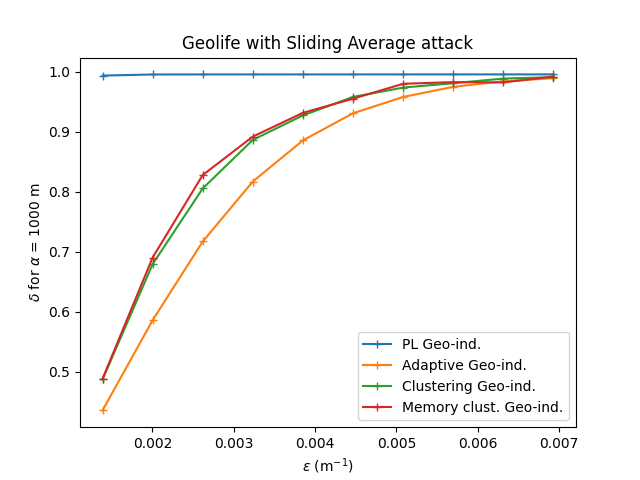}
    \caption{($\alpha$, $\delta$)-usefulness for $\alpha$ = 1~km on Geolife dataset with sliding average attack}
    \label{fig:usefulness_1000_geolife_de_obf_sliding_metric_alpha_delta}
\end{figure}

\FloatBarrier

\subsection{Gowalla}
\begin{figure}[h!]
    \centering
    \includegraphics[width=.8\linewidth]{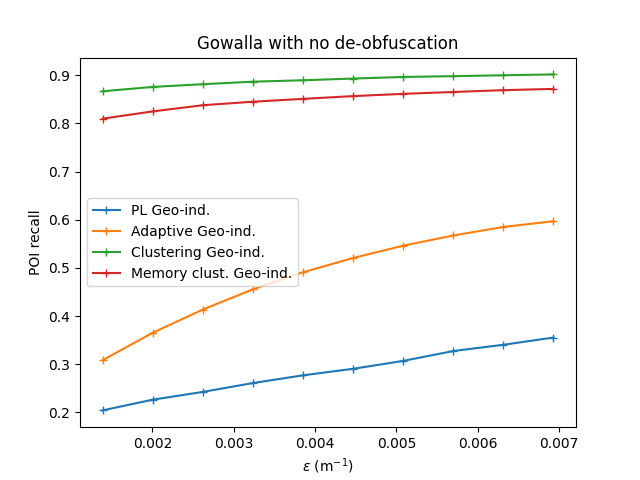}
    \caption{POI recall on Gowalla dataset}
    \label{fig:gowalla_metric_POIMetric}
\end{figure}
\begin{figure}[h!]
    \centering
    \includegraphics[width=.8\linewidth]{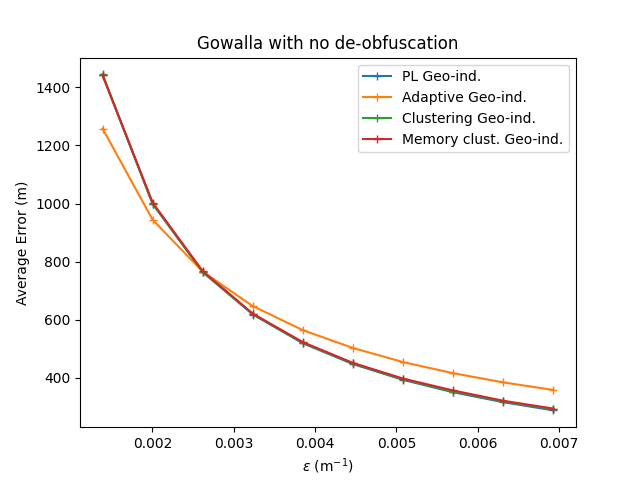}
    \caption{Average error on Gowalla dataset}
    \label{fig:gowalla_metric_avg_error2}
\end{figure}
\begin{figure}[h!]
    \centering
    \includegraphics[width=.8\linewidth]{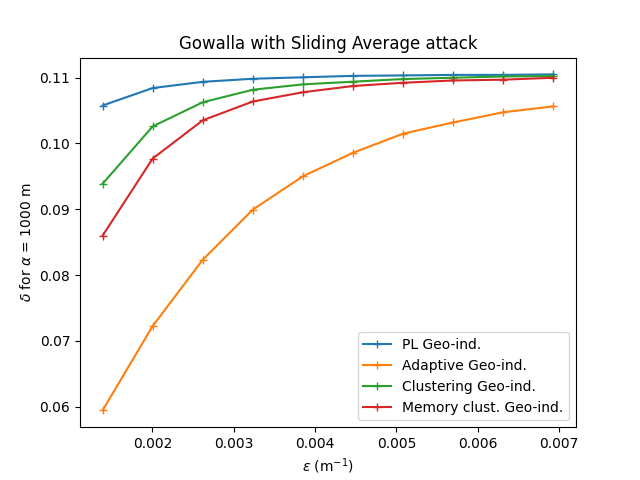}
    \caption{($\alpha$, $\delta$)-usefulness for $\alpha$ = 1~km on Gowalla dataset with sliding average attack}
    \label{fig:usefulness_1000_gowalla_de_obf_sliding_metric_alpha_delta}
\end{figure}

\FloatBarrier

\subsection{San Francisco Cabs}
\begin{figure}[h!]
    \centering
    \includegraphics[width=.8\linewidth]{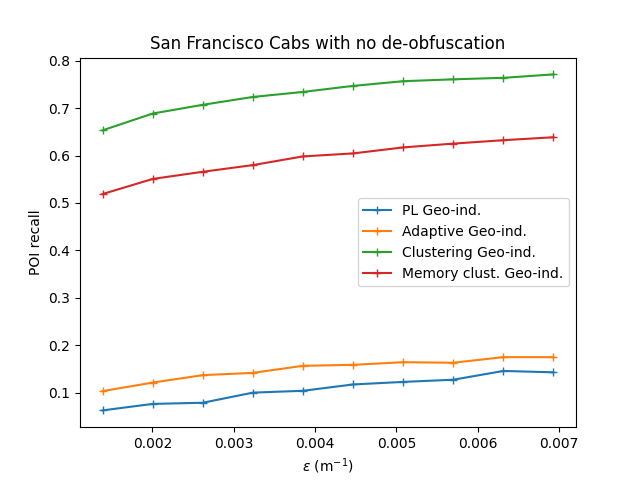}
    \caption{POI recall on San Francisco Cabs dataset}
    \label{fig:sf_cabs_metric_POIMetric}
\end{figure}
\begin{figure}[h!]
    \centering
    \includegraphics[width=.8\linewidth]{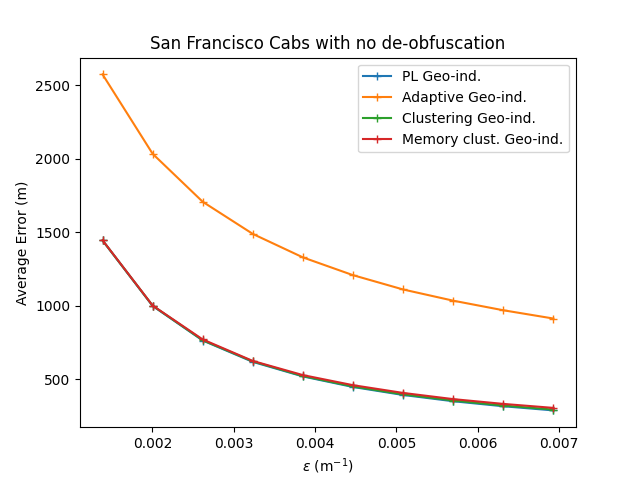}
    \caption{Average error on San Francisco Cabs dataset}
    \label{fig:sf_cabs_metric_avg_error2}
\end{figure}
\begin{figure}[h!]
    \centering
    \includegraphics[width=.8\linewidth]{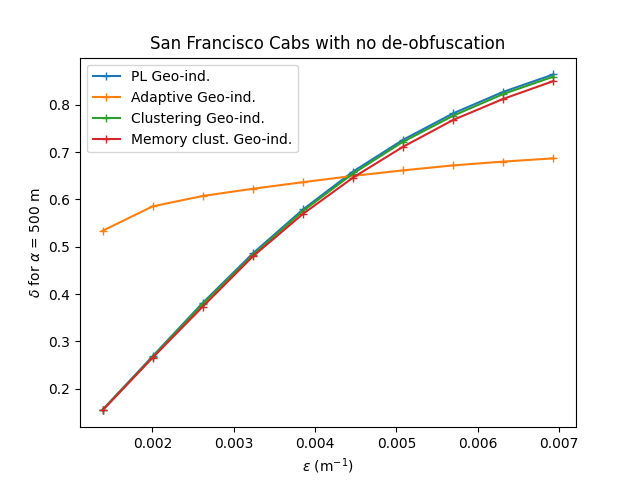}
    \caption{($\alpha$, $\delta$)-usefulness for $\alpha$ = 500~m on San Francisco Cabs dataset}
    \label{fig:usefulness_500_sf_cabs_metric_alpha_delta}
\end{figure}

\FloatBarrier

\subsection{Temporally sub-sampled dataset}
\begin{figure}[h!]
    \centering
    \includegraphics[width=.8\linewidth]{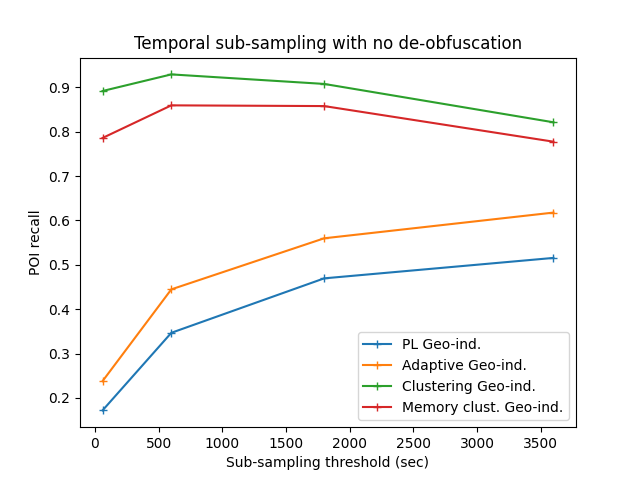}
    \caption{POI recall on temporally sub-sampled datasets}
    \label{fig:temp_sample_metric_POIMetric2}
\end{figure}
\begin{figure}[h!]
    \centering
    \includegraphics[width=.8\linewidth]{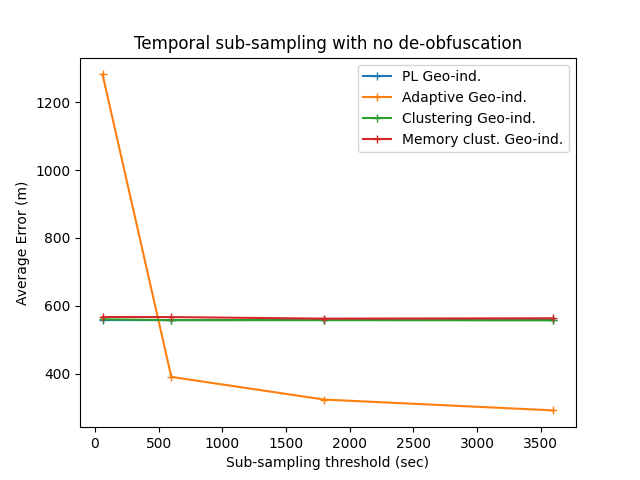}
    \caption{Average error on temporally sub-sampled datasets}
    \label{fig:temp_sample_metric_avg_error}
\end{figure}
\begin{figure}[h!]
    \centering
    \includegraphics[width=.8\linewidth]{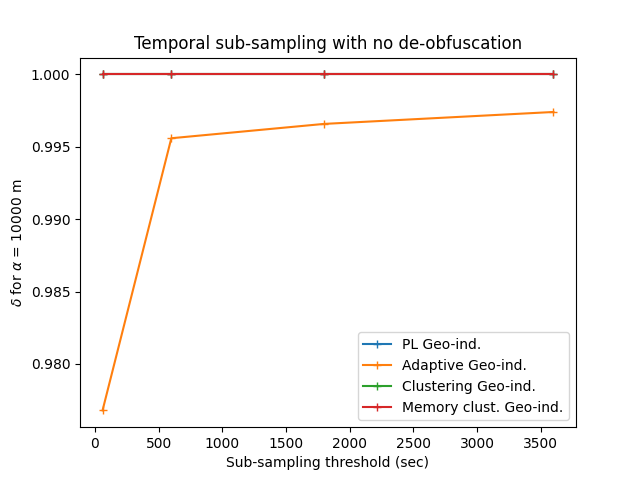}
    \caption{($\alpha$, $\delta$)-usefulness for $\alpha$ = 10~km on temporally sub-sampled datasets}
    \label{fig:usefulness_10000_temp_sample_metric_alpha_delta}
\end{figure}
\begin{figure}[h!]
    \centering
    \includegraphics[width=.8\linewidth]{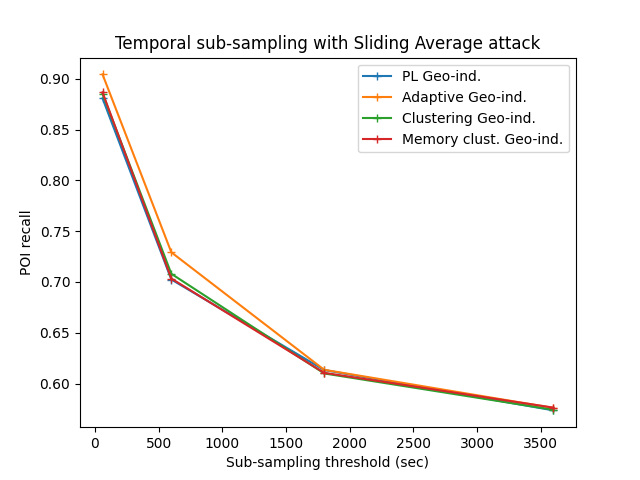}
    \caption{POI recall on temporally sub-sampled datasets with sliding average attack}
    \label{fig:temp_sample_de_obf_sliding_metric_POIMetric}
\end{figure}
\begin{figure}[h!]
    \centering
    \includegraphics[width=.8\linewidth]{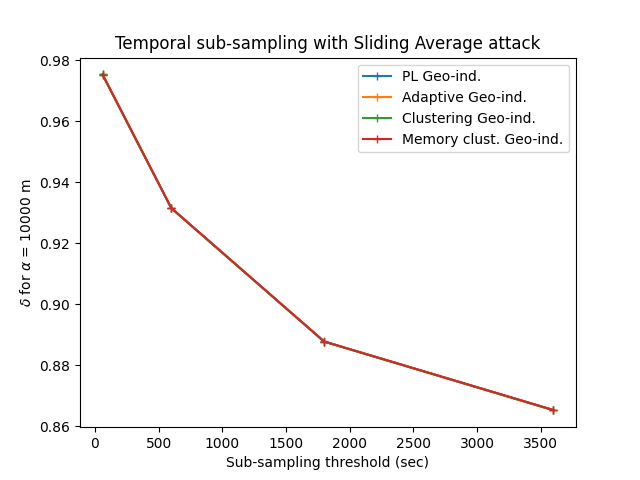}
    \caption{($\alpha$, $\delta$)-usefulness for $\alpha$ = 10~km on temporally sub-sampled datasets with sliding average attack}
    \label{fig:usefulness_10000_temp_sample_de_obf_sliding_metric_alpha_delta}
\end{figure}

\FloatBarrier

\subsection{Spatially sub-sampled dataset}
\begin{figure}[h!]
    \centering
    \includegraphics[width=.8\linewidth]{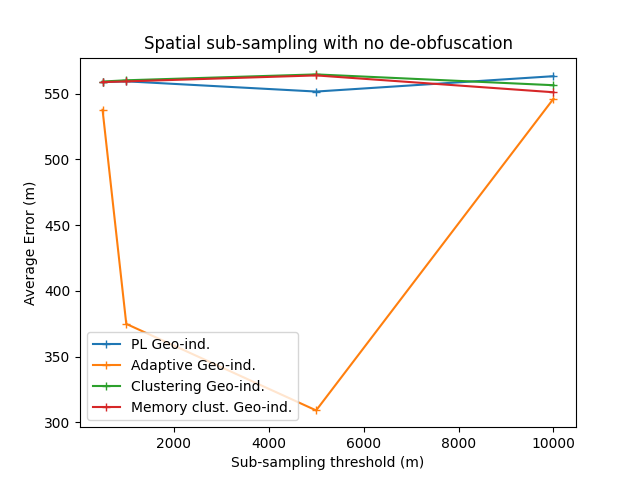}
    \caption{Average error on spatially sub-sampled datasets}
    \label{fig:spatial_sample_metric_avg_error}
\end{figure}
\begin{figure}[h!]
    \centering
    \includegraphics[width=.8\linewidth]{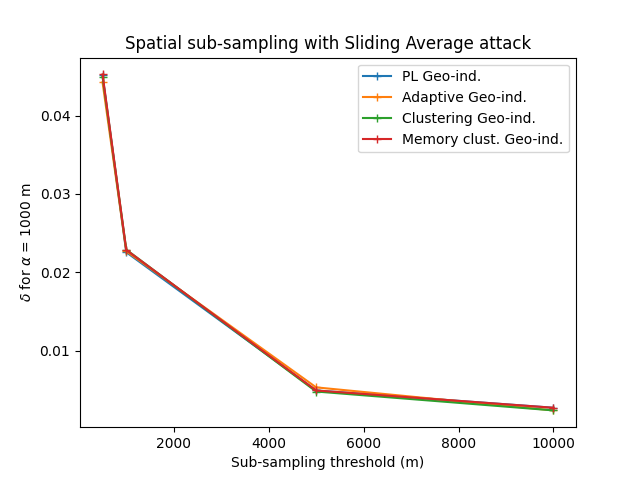}
    \caption{($\alpha$, $\delta$)-usefulness for $\alpha$ = 1~km on spatially sub-sampled datasets}
    \label{fig:usefulness_1000_spatial_sample_de_obf_sliding_metric_alpha_delta}
\end{figure}
\begin{figure}[h!]
    \centering
    \includegraphics[width=.8\linewidth]{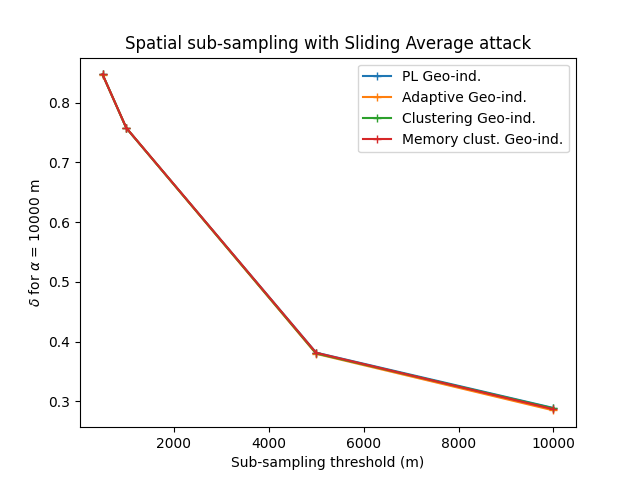}
    \caption{($\alpha$, $\delta$)-usefulness for $\alpha$ = 10~km on spatially sub-sampled datasets with sliding average attack}
    \label{fig:usefulness_10000_spatial_sample_de_obf_sliding_metric_alpha_delta}
\end{figure}